\documentclass[reprint,prl,superscriptaddress]{revtex4-1}
\usepackage{bm}
\usepackage{color}

\newcommand\nhat{\hat{\mathbf n}}

\newcommand\beq{\begin{equation}}
\newcommand\eeq{\end{equation}}
\newcommand\beqn{\begin{eqnarray}}
\newcommand\eeqn{\end{eqnarray}}
\newcommand\ave[1]{\left\langle {#1} \right\rangle}

\newcommand\cl{C_{\ell}}

\newcommand\fsky{f_{\mathrm{sky}}}
\newcommand\nn{\nonumber}

\def\cl{C_{\ell}}

\def\Om{\ensuremath{\Omega_{\mathrm{m}}}}

\def\2gcm{\textrm{g cm$^{-2}$}}

\def\H0{\ensuremath{\mathrm{H}_0}}
\def\nhat{\hat{\mathbf{n}}}

\def\nn{\nonumber}

\def\fsky{f_{\mathrm{sky}}}

\def\gcmb{\ensuremath{g_{\mathrm{\scriptscriptstyle{CMB}}}}}
\def\ggal{\ensuremath{g_{\mathrm{\scriptstyle opt}}}}
\def\ngal{\ensuremath{N_{\mathrm{\scriptstyle opt}}}}
\def\nfg{\ensuremath{N_{\mathrm{\scriptstyle f}}}}
\def\kcmb{\ensuremath{\kappa_{\mathrm{\scriptscriptstyle{{CMB}}}}}}
\def\kgal{\ensuremath{\kappa_{\mathrm{\scriptscriptstyle{opt}}}}}

\def\nl{N_\ell}

\def\be{\begin{equation}}
\def\ee{\end{equaiton}}

\usepackage{graphicx}
\usepackage[large]{subfigure}
\usepackage{amssymb, amsmath}
\usepackage[amssymb]{SIunits}
\usepackage{hyperref}
\usepackage{aas_macros}
\usepackage{natbib}
\renewcommand\section[1]{\emph{#1}.---}

\begin{document}
\title{Can CMB Lensing Help Cosmic Shear Surveys? }
\author{Sudeep Das}
\affiliation{High Energy Physics Division, Argonne National Laboratory, 9700 S Cass Avenue, Lemont, IL 60439}
\affiliation{Berkeley Center for Cosmological Physics, Berkeley, CA 94720}
\author{Josquin Errard}
\affiliation{Computational Cosmology Center, Lawrence Berkeley National Laboratory, Berkeley, CA 94720}
\author{David Spergel}
\affiliation{Peyton Hall, Ivy Lane, Princeton University, Princeton, NJ 08544}
%\author{ others} 
\date{\today}                                           % Activate to display a given date or no date
\begin{abstract}{Yes! Upcoming galaxy shear surveys have the potential to significantly improve our understanding of 
dark energy and neutrino mass {\it if} lensing systematics can be sufficiently controlled. The cross-correlations between the weak lensing shear,  galaxy number counts from a galaxy redshift survey, and the  CMB lensing convergence can be used to calibrate the shear multiplicative bias, one of the most challenging systematics in lensing surveys. These cross-correlations can significantly reduce the deleterious effects of the uncertainties in multiplicative bias. }
\end{abstract}
\maketitle
%\section{}
%\subsection{}
\section{Introduction} 
The large scale structure in the universe gravitationally deflects the light from distant  galaxies inducing 
weak coherent distortions  of their images.  This ``cosmic shear" signal \cite{hoekstra_jain_2008,refregier_2003}, depends sensitively on the expansion rate of the universe as well the growth of structure with time, making it a  potentially rich probe of the nature of dark energy, the validity of general relativity (GR) on cosmological scales,  and the neutrino mass sum~\cite{1987tasi.rept.....A,2006Msngr.125...48P,2010MNRAS.404..110D,1999ApJ...514L..65H,2002PhRvD..65f3001H,2013arXiv1301.1037D,2013arXiv1309.5383A}.  In the next few years, we expect a flood of data relevant to cosmic shear studies from ongoing, upcoming and planned galaxy surveys, such as PanSTARRS~\footnote{http://pan-starrs.ifa.hawaii.edu/},  the Subaru HyperSuprimeCam (HSC) survey~\footnote{http://sumire.ipmu.jp/en/3358}, the Dark Energy Survey (DES)~\footnote{http://www.darkenergysurvey.org/survey/},  KIDS~\footnote{http://kids.strw.leidenuniv.nl}, 
LSST~\footnote{www.lsst.org}, Euclid~\footnote{http://www.euclid-ec.org/},~WFIRST~\footnote{http://wfirst.gsfc.nasa.gov}, BigBoss~\footnote{http://bigboss.lbl.gov/}, etc. The weak lensing  signal from these surveys can in principle pin down cosmological parameters to  unprecedented  precision, but realizing their full potential imposes stringent requirements on the control of systematics~\cite{2008MNRAS.391..228A, 2006MNRAS.366..101H}.  \par
Weak lensing systematics mainly stem from four sources: the inability to accurately measure galaxy shapes due to 
instrumental and atmospheric effects, the uncertainty in the distance to the background galaxies (photometric redshift errors),  intrinsic alignment of galaxies due to  the galaxy formation process, and 
the theoretical uncertainties in dark matter clustering on small scales. Shape measurement errors fall mainly 
into two categories -- additive  and   multiplicative biases in the deduced shear signal, both of which can be 
redshift dependent in general. The multiplicative bias (which multiplies  the true shear signal with an unknown multiplicative factor) is particularly notorious because being redshift dependent it can be degenerate with the growth
of structure,  significantly degrading cosmological parameter constraints and inducing large parameter biases~\cite{2008MNRAS.391..228A, 2006MNRAS.366..101H}.  \par
A multiplicative bias in the shear signal can in principle be calibrated  by cross correlating the observed shear signal with  a measurement of the projected dark matter field that does not suffer from such multiplicative uncertainty. The gravitational lensing of the cosmic microwave background (CMB) provides a promising route to such a solution, as we describe in this Letter.  Recently, Vallinotto (2012)~\cite{2012ApJ...759...32V} has proposed using the 
cross-correlation of CMB lensing with cosmic shear as a method to control the multiplicative bias. It is an useful 
method for controlling a redshift dependent multiplicative bias,  as long as the growth function is assumed to be standard. However, a non-standard growth function can be hidden by a redshift dependent multiplicative bias even when using this cross-correlation technique. We propose that by additionally using cross-correlations with spectroscopic galaxy surveys such  degeneracy can be further broken, significantly improving the constraints on dark energy, deviation from GR, and neutrino mass. \\

\section{Controlling Shear Multiplicative Bias with Cross-correlations}
We begin by writing the weak lensing convergence as the projected  matter overdensity field $\delta$:
\be
\kappa(\nhat)=\frac32 \Om H_{0}^{2}\int  d\eta ~d_{A}^{2} (\eta) \frac{g(\eta)}{a(\eta)}\delta(d_A(\eta) \nhat,\eta),
\eeq 
where the kernel
\beq
g(\eta)=\frac{1}{d_A(\eta)}\int_{\eta}^{\infty}d\eta' ~W_b(\eta') \frac{d_A(\eta'-\eta)}{d_A(\eta') }
\eeq
depends on the normalized  source distribution in comoving distance $\eta$: $W_b(\eta)$.  Here 
 $d_A(\eta)$ is  the comoving angular diameter distance,  $a(\eta)$ is the scale factor, while $\Om$  and $H_{0}$ represent the present values of the matter density parameter and the Hubble parameter, respectively. For the CMB we can approximate the source distribution function as a screen at the  last scattering distance $\eta_0$:  $W_{b}(\eta) \simeq \delta_D(\eta - \eta_{0})$.
\par 
Next we consider a  tracer population  with a known redshift distribution  $W_f(\eta)$: 
\beq
\Sigma(\nhat)=\int d\eta  ~W_f(\eta) \delta_g(\eta \nhat,\eta),
\eeq  
where $\delta_g$  represents the fractional tracer overdensity that is related to the underlying matter density field via a scale and redshift dependent bias factor:  $\delta_{g}(\mathbf k,\eta) = b(k,\eta) \delta(\mathbf k,\eta)$.
CMB lensing estimators lets us reconstruct the convergence field, $\kappa_{CMB}$ between us and the last scattering surface. 
The observed CMB lensing-tracer cross-correlation depends on the bias of the galaxy distribution, $b_\ell(\eta)$ and
the underlying cosmology,
\beq
\cl^{\kcmb\Sigma}=\frac{3}{2}{\Om H_{0}^{2}}  \int d\eta ~b_\ell(\eta) W_f(\eta) \frac{\gcmb(\eta)}{a(\eta)}  P\left(\frac{\ell}{d_A},\eta\right),
\eeq
where we have used the Limber approximation, the orthogonality of spherical harmonics,  and employed the shorthand notation, $b_\ell(\eta) \equiv  b(\frac{\ell}{d_A},\eta)$.\par 
The weak lensing-tracer cross-correlation is similarly computed, but it  depends  also on the shear multiplicative bias, $m$:
\beq
\cl^{\kgal\Sigma}= m ~ \frac{3}{2}{\Om H_{0}^{2}}  \int d\eta   ~ b_\ell(\eta) W_f(\eta) \frac{\ggal(\eta)}{a(\eta)}  P\left(\frac{\ell}{d_A},\eta\right). 
\eeq
Note that if the tracer redshift distribution is narrow, then the ratio between the two reduces to a product of the multiplicative bias times the geometric distance
ratio:
\begin{equation}
 \frac{\cl^{\kgal \Sigma}} {\cl^{\kcmb\Sigma}}= m\frac{\ggal(\eta)}{\gcmb(\eta)}
 \end{equation}
 
 As Vallinotto (2012)~\cite{2012ApJ...759...32V} noted, we can also estimate this multiplicative bias by looking at the ratio of the convergence power spectra:
 \begin{equation}
\frac{\cl^{\kcmb\kgal}}{\cl^{\kcmb\kcmb}} =  m \frac{\int d\eta\ d_A^2(\eta)\frac{\gcmb(\eta) \ggal(\eta)}{a^2(\eta)} P(\frac{\ell}{d_A},\eta)}{\int d\eta\ d_A^2(\eta)\frac{\gcmb^2(\eta)}{a^2(\eta)} P(\frac{\ell}{d_A},\eta)}\end{equation}
or $\cl^{\kcmb\kgal}/{\cl^{\kgal\kgal}}$ which has a different dependence on cosmological parameters than the tracer-lensing ratio.\\

\begin{figure*}
\includegraphics[width=\columnwidth]{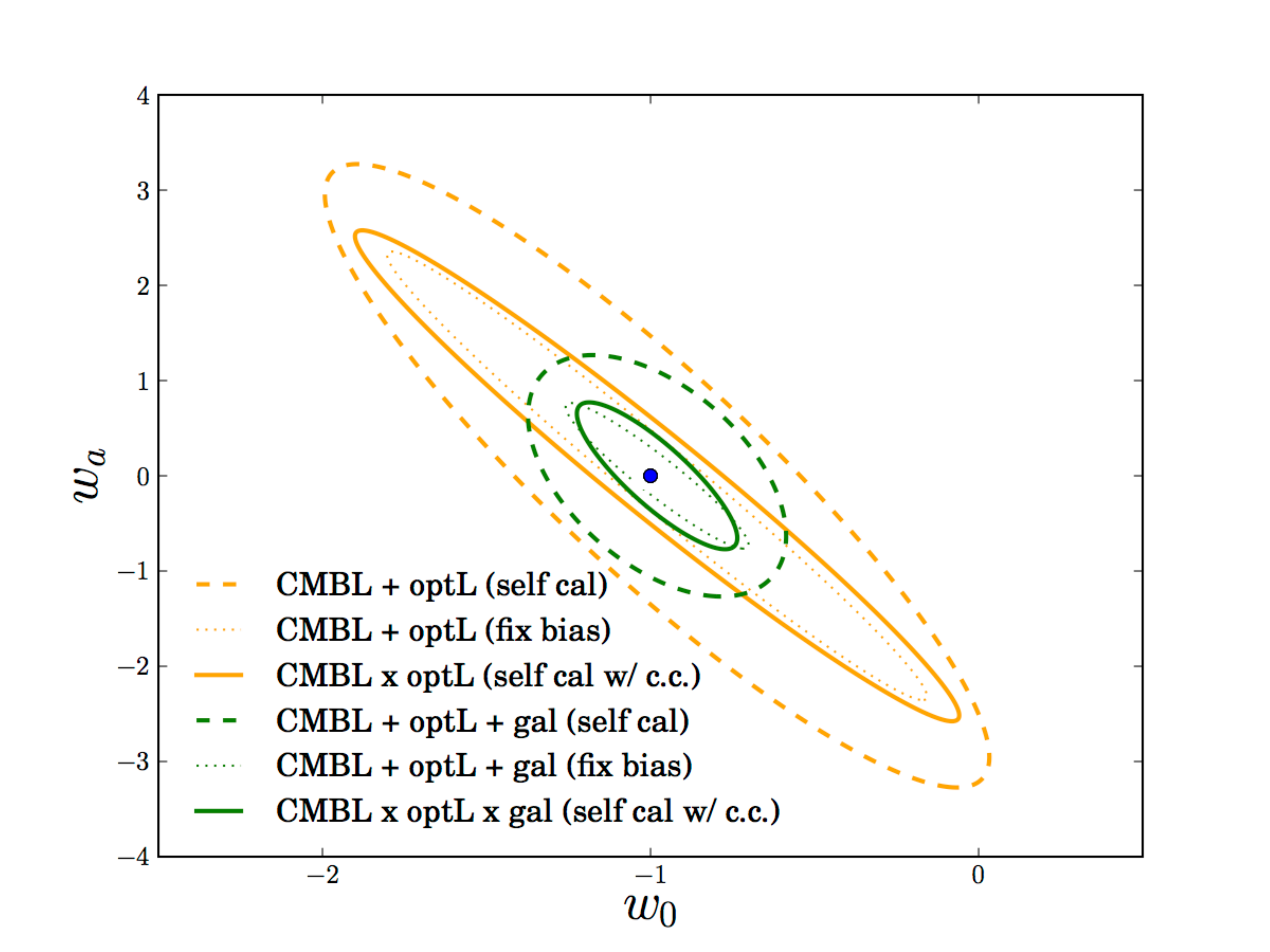}
\includegraphics[width=\columnwidth]{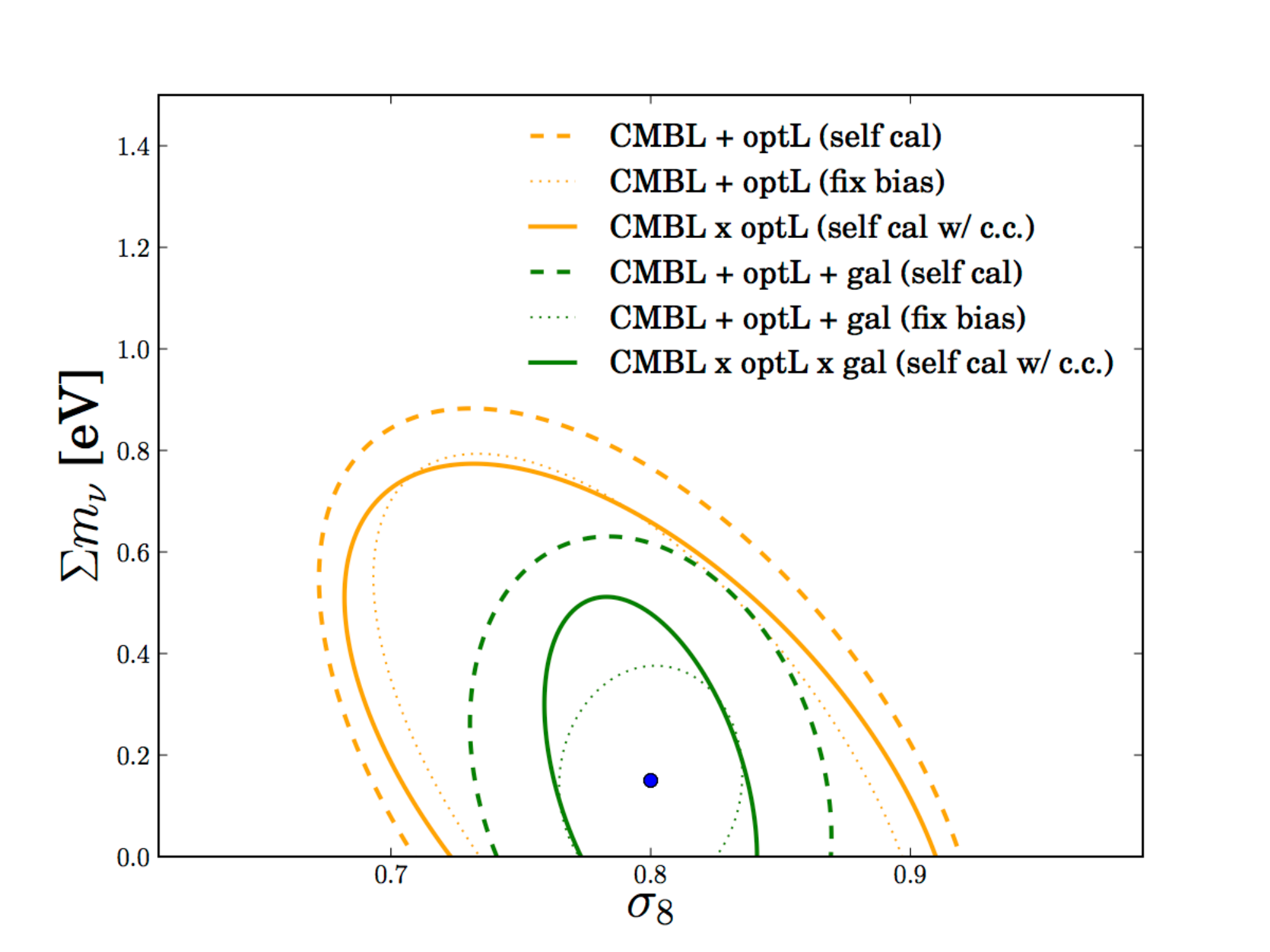}
\caption{\emph{Left:} Marginalized 68\% confidence contours on the dark energy equation of state parameters using various combinations of observables.  The outer (orange)  dashed ellipse is from the addition of the deflection field power spectrum from CMB lensing reconstruction to the tomographic shear  power spectra from optical lensing, when the shear multiplicative 
biases are marginalized over along with other cosmological parameters, including neutrino mass sum. The outer (orange) dotted ellipse is the same as above, but with the biases held fixed at their fiducial values.  The outer solid (orange) ellipse results from the addition of the CMB lensing - optical lensing cross-correlations to the above, with the multiplicative biases marginalized over (and thereby self-calibrated).  The inner (green)  dashed contour results from adding, without any cross-correlation, the CMB lensing, the optical lensing, and the galaxy angular clustering measurements with both the shear biases and the galaxy linear biases marginalized over. The inner (green) dotted contour results from holding these biases fixed, while the inner (solid) ellipse is the result of including cross-correlations between CMB lensing, optical lensing, and galaxy angular clustering, with all the biases marginalized over. The resulting constraints on the biases are displayed in Table~\ref{biasTable}.  \emph{Right:} The interpretation is the same as left panel, except here we show the constraints on the neutrino mass sum and $\sigma_8$. Note that in this case, uncertainties in $w_0$, $w_a$ are always marginalized over. Complementary constraints are summarized in Table ~\ref{paramsTable}. See text for survey details.  \label{ellipses}}
\end{figure*}

\section{A Fisher Matrix Approach} We study the effectiveness of the cross-correlation statistics in controlling systematics and their effects on  cosmological parameters using a Fisher Matrix approach.  We assume $\ngal$  tomographic redshift bins for the weak lensing survey and $\nfg$ redshift slices constructed from the photometric galaxy survey. The data vector consists of the possible auto and cross spectra between weak lensing, galaxy  angular clustering, and CMB lensing: 
\beqn
\nn C_\ell^{XY} &\equiv& \left\{\cl^{\kcmb\kcmb}, \underbrace{\{\cl^{\kcmb\kgal}\}}_{\ngal\ \mathrm{spectra}},  \underbrace{ \{\cl^{\kcmb\Sigma}\}}_{\nfg\ \mathrm{spectra}}, \right.\\
&&\nn \left. \underbrace{\{ \cl^{\kgal\kgal}\}}_{\ngal (\ngal +1)/2\ \mathrm{spectra}}, \underbrace{\{\cl^{\kgal\Sigma}\}}_{\ngal \nfg\  \mathrm{spectra}},\right. \\ 
&& \left. \underbrace{\{\cl^{\Sigma\Sigma}\}}_{\nfg (\nfg+1)/2\ \mathrm{spectra}} \right\} . 
\eeqn
The Fisher Matrix is constructed as, 
\beq
F_{\alpha\beta} = \sum_\ell  \sum_{\substack{ \left\{X, Y, W, Z \right\} \in \\ \left\{ \kcmb, \kgal, \Sigma \right\}}} \left[ \frac{\partial C_\ell^{XY}}{\partial \alpha}\left( \mathbf{Cov}_\ell^{{XY, WZ}}\right)^{-1}\frac{ \partial C_\ell^{WZ}}{\partial \beta}\right]
\eeq
where $\alpha, \beta$ run over cosmological as well as nuisance parameters (e.g. multiplicative biases), and the covariance matrix is computed under the assumption of Gaussian random fields: 
\beq
 \mathbf{Cov}_\ell^{{XY, WZ}} = \frac{1}{(2\ell+1) \fsky}  \left( \tilde \cl^{XW} \tilde\cl^{YZ} + \tilde \cl^{XZ} \tilde\cl^{YW}  \right)
\eeq
 where $\fsky$ is the fraction of sky covered by the overlapping experiments, and $\tilde \cl \equiv \cl+ \nl$ includes the cosmological signal and noise.  The noise for the CMB lensing reconstruction $\nl^{\kcmb}$ is computed 
 using the quadratic estimator technique \cite{hu/okamoto:2002}. Weak lensing noise in a tomographic bin gets contributions from the shape noise $\nl^{\kappa_{opt}} = {\ave{\gamma^2}^{1/2}/\bar n}$ and additive shear errors, following the implementation described in~\cite{2013arXiv1301.1037D}.  Lastly, we assume that the noise in the angular correlation of tracers in a redshift bin is purely shot noise dominated:  $N^{\Sigma}_\ell= {1}/{\bar n_f}$, where $\bar n_f$ is the number of tracers per steradian in that bin. \\
 
\section{Surveys}
For a near term application of our proposed method we consider the cross-correlations between the weak lensing signal from a HSC-like survey, the CMB lensing from an ACTPol-like experiment and galaxies from a BOSS-like survey. We assume that HSC, the CMB ground based, and the BOSS-like  surveys have the same footprint of 4000 deg$^2$ on the sky.  For the ACTPol-like survey, we assume a depth of 5 (7) $\mu$K-arcmin in temperature (polarization) and a $1.4$ arcmin beam.  The HSC weak lensing sources are assumed to be distributed in redshift as 
\begin{eqnarray}
	\centering
		n(z) = \frac{3 N_g}{2  z_0^3 } \ z^2 \exp\left[-\left(\frac{z}{z_0}\right)^{3/2}\right]
\end{eqnarray}
with $z_0 = 0.69$ corresponding to a median redshift of unity, and  a source density of $N_g$ =  35 galaxies per arcmin$^2$. We assume a photometric redshift error distribution of $\sigma(z) = 0.03 (1+z)$, and an intrinsic shape noise of $\langle \gamma^2\rangle ^{1/2} = 0.4$.  We consider three tomographic bins from $0< z <0.7$, $0.7<z<1.5$, $1.5<z<4.0$. We treat the multiplicative bias $m = 1+ \alpha$ in each bin as an independent parameter with empirically motivated  fiducial values of $\alpha$=  0.008,  0.01, and  0.02.  As the focus of the study is the multiplicative bias, we do not treat the additive shear errors for reasons of clarity.  For the BOSS-like survey, we assume three redshift bins $0.3<z<0.4$, $0.4<z<0.5$ and $0.5<z<0.6$ with the linear bias parameter of 2.0 in each bin (which are marginalized over), and a total galaxy density of $0.011$ per arcmin$^2$. \par 

For the Fisher analysis, we consider the standard set of  $\Lambda$CDM parameters with fiducial values: $\Omega_b h^2= 0.02258$, $\Omega_c h^2 = 0.1093$, $\Omega_\Lambda = 0.734$, $n_s = 0.96$ and $\sigma_8 = 0.8$, plus a massive neutrino component $\Omega_\nu h^2 = 0.001596$ corresponding to $\sum m_\nu = 0.15$ eV, and the  dark energy equation of state parameterized via   $w(z) = w_0 + w_a z/ (1+z) $.  We vary these parameters as well as 
the tracer galaxy biases in each spectroscopic bin, and the shear multiplicative biases in each tomographic bin. 
We do not include any CMB power spectra information in this analysis.\\

\section{Results} 
We first confine ourselves to  the optical lensing and CMB lensing convergence fields only, and consider only the internal spectra so that the data vector only consists of  $\{\cl^{\kcmb\kcmb}$,   $\cl^{{\kgal}_i {\kgal}_j}\}$.  We perform two parameter estimation 
studies: one with the multiplicative biases fixed at their fiducial value  [CMBL+ optL (fix bias)] and the other letting these biases vary and be thereby self-calibrated [CMBL+ optL (self cal)].
Note that although we do not consider the cross-correlation between CMB lensing and optical lensing as an observable yet, such cross-correlation terms do appear in  the covariance  matrix. We cannot simply add the CMB lensing and optical lensing Fisher matrices here, as the two effects are correlated.  Fig.~\ref{ellipses} and Table~\ref{paramsTable} show the resulting constraints on the dark energy equation of state and the neutrino mass sum for these two scenarios (note that we always marginalize over all other parameters in these plots, and complementary cases are summarized in the table). Letting the multiplicative shear float significantly degrades both the dark energy and neutrino figures of merit. Next, we expand our data vector to include the CMB lensing-optical lensing cross-correlation: $\{\cl^{\kcmb\kcmb}$,   $\cl^{{\kgal}_i {\kgal}_j}$, $\cl^{{\kcmb}_i {\kgal}_j}\}$ and let the biases vary  [CMBL $\times$ optL (self cal with c.c.)].  This shows the power of the cross-correlations ---  including the cross-correlation leads to  self calibration of  the shear multiplicative biases to  sufficient accuracy so that the parameter uncertainty ellipses shrink to give almost the same figures of merit as the case where the multiplicative biases were held fixed.  The corresponding marginalized constraints on the multiplicative biases are shown in Table~\ref{biasTable}. Note that $\sim$10\%  deviations of the multiplicative biases from unity 
can be calibrated by this method. The constraints will improve significantly in future with larger  and deeper weak lensing/CMB lensing surveys, e.g.~\cite{2013arXiv1309.5383A}. \par
Next, we include   the three BOSS-like galaxy number density bins in the analysis. 
 First, we look at the constraints avoiding any cross-correlations , once fixing both shear and galaxy biases [CMBL + optL + gal (fix bias)], and then letting the biases self calibrate [CMBL + optL + gal (self cal)].  As in the previous case, letting biases float significantly degrades all constraints. There are two interesting things to note here: first, just adding galaxy clustering  information significantly improves dark energy and neutrino mass. Second, and related to the above,  just by having these overlapping data sets self calibrates the galaxy biases to 3-4\% This is not surprising because both 
 CMB lensing and optical lensing constrains growth (and $\sigma_8$) and  galaxy biases. For comparison, if we only had the BOSS-like survey, then the Fisher analysis predicts galaxy bias errors of $(0.23, 0.24, 0.28 )$.  Next we introduce all possible cross-correlations between the  CMB lensing,  optical lensing, and the galaxy density fields [CMBL $\times$ optL $\times$ gal (self cal w/ c. c.)]. We observe a significant improvement the shear multiplicative bias constraints in this case, with the deviation from unity constrained down 
 to $0.5\%$ in the highest redshift bin.  We also notice an appreciable improvement in the constraints on the galaxies bias parameters. These improvements propagate to cosmological constraints, and one can get significantly closer through this cross-correlation procedure to the dark energy (neutrino)  figures of merit where the biases are 
 artificially held fixed. Having a spectroscopic sample breaks the degeneracy between multiplicative bias and growth that limits the use of lensing correlations alone \cite{2012ApJ...759...32V}.
 
\begin{table} 
\caption{Estimated marginalized 1-$\sigma$ error on the shear multiplicative bias parameter $m_i= (1+\alpha_i)$ in the tomographic bin $i$, 
 and the galaxy  bias parameter $b_j$ in spectroscopic bin $j$ for various ways of combining data sets.\label{biasTable}}
\center
\begin{tabular}{l|l|l|l|l|l}
\hline
\hline
Bias & Fiducial & CMBL  & CMBL &  CMBL  &  CMBL \\
parameter & value & + optL & $\times$ optL & + optL  & $\times $ optL \\
& & & & + gal. & $\times $ gal \\
\hline
\hline
$\alpha_1$ & 0.008  & 0.058 & 0.026 & 0.047 & 0.021\\ 
$\alpha_2$ & 0.014  & 0.063 & 0.010 & 0.054 & 0.008\\ 
$\alpha_3$ & 0.020  & 0.063 & 0.007 & 0.053 & 0.005\\ 
\hline
$b_1$ &  2.000 & - & - & 0.070 & 0.053 \\ 
$b_2$ &  2.000 & - & - & 0.058 & 0.044 \\ 
$b_3$ &  2.000 & - & - & 0.073 & 0.055 \\
\hline
\hline
\end{tabular}
\end{table}

\begin{table*} 
\caption{Estimated 1-$\sigma$ error on chosen cosmological parameters: $\left\{ \sigma_8,  \Sigma m_\nu, w_0, w_a \right\}$ \emph{marginalized} over the usual $\Lambda$CDM parameters. Numbers with (without) parenthesis are obtained with fixed (varying) bias parameters. A dash means that the constraint has not been marginalized over this parameters\label{paramsTable}.}
\center
{\tiny
\begin{tabular}{||c|c|ccc|ccc|ccc|ccc||}
\hline
\hline
& \textbf{Fiducial} &  \multicolumn{3}{c|}{\textbf{CMBL}}  & \multicolumn{3}{c|}{\textbf{CMBL}} &  \multicolumn{3}{c}{\textbf{CMBL}}  &  \multicolumn{3}{|c||}{\textbf{CMBL}}  \\
 & \textbf{value} & \multicolumn{3}{c}{\textbf{+ optL} }& \multicolumn{3}{|c}{$\times$ \textbf{optL}} & \multicolumn{3}{|c}{\textbf{+ optL}}  & \multicolumn{3}{|c||}{$\times $ \textbf{optL}} \\
 &  & \multicolumn{3}{c|}{ }& \multicolumn{3}{c|}{ } &  \multicolumn{3}{c|}{\textbf{+ gal.}} & \multicolumn{3}{c||}{$\times $ \textbf{gal}} \\
\hline
\hline
$\sigma \left( \sigma_8 \right)$ &                            0.80 & 0.0211  & 0.0524 & 0.0842 &0.0172 & 0.0238   & 0.0777  &  0.0199   & 0.0470   & 0.0642 & 0.0158  & 0.0203 & 0.0498 \\ 
 &    & (0.0144) & (0.0160) &  (0.0704) &  (0.0133) &  (0.0149) &   (0.0683) &  (0.00484) & (0.0114)   &  (0.0381) & (0.0112) & (0.0120) & (0.0464) \\ 
\hline
$\sigma\left( \Sigma m_\nu \right) [{\rm meV}]$ & 150  & 305  & 327 & 483 & 208 & 311 & 411 & 273 & 288  & 414 & 191   & 257 & 350 \\ 
& &   (156)  &  (313) & (424) & (142) & (275)  & (380) & (128) & (184) & (199)   & (138) & (240) & (327)  \\ 
\hline
$\sigma\left( w_0 \right)$ & -1.0  & -  & 0.288 & 0.668  & -    & 0.132 & 0.607 & -  & 0.256 & 0.489 &  -  & 0.103  & 0.381 \\  
 & & & (0.0983)   & (0.542)     &  & (0.0921)  & (0.523)    &  & (0.0677) & (0.346)  &  & (0.0779) & (0.359) \\  
\hline
$\sigma\left( w_a \right)$ & 0.0  & - &-   & 2.16 & - &   -   & 1.70 & -  & -  & 1.63  & -    & -   & 1.14 \\ 
& & & & (1.55) & & & (1.49)  &  & & (0.932)  & & & (1.06) \\ 
\hline
\hline
\end{tabular}}
\end{table*}

\section{Conclusions}
The world astrophysics community is about to embark on several major weak lensing surveys.  Over the coming decade, we are likely to invest several billion dollars in missions among whose primary goals is to use weak lensing
measurements to determine the nature of dark energy, the geometry of the universe and to constrain the nature of dark matter.  Multiplicative biases may well limit the scientific utility of these surveys.
In this letter, we show that by combining optical weak lensing measurements with CMB lensing measurements and a spectroscopic survey, we can calibrate the amplitude of the multiplicative bias in the optical lensing survey.
This will likely noticeably improve our ability to measure cosmological parameters. In particular, it is remarkable that the Fisher cross terms brings enough information to significantly shrink uncertainties on parameters estimation.

For the upcoming HSC survey, the combination of CMB lensing measurements from an ACTPol-like experiment and BOSS galaxy spectroscopy should enable a calibration of the multiplicative bias at the $0.5-2.0\%$ level, and reduce the error bar on cosmological parameters by $\sim 20-25\%$.
In the near future, stage-III CMB experiments should be able to make accurate enough CMB lensing measurements to be able to calibrate optical weak lensing surveys like LSST at the $0.25\%$ level.

\acknowledgments{The Berkeley Weak Lensing code developed by Sudeep Das, Reiko Nakajima, Roland de Putter, and Eric Linder has been used extensively in this study.  We thank Gil Holder, Alberto Vallinotto and Benjamin Joachimi for useful comments and discussions.  }

\bibliography{act_lensing}

\end{document}